\documentclass[traditabstract,rnote]{aa}

\usepackage{txfonts}
\usepackage{natbib}
\usepackage{graphicx}
\usepackage{mathrsfs}

\bibpunct{(}{)}{;}{a}{}{,}

\newcommand{\targetnts}{OGLE052218.07-692827.4}
\newcommand{\target}{OGLE052218.07-692827.4\ }
\newcommand{\tcen}[1]{\multicolumn{1}{c}{#1}}

\begin{document}

\title{In search of RR Lyrae type stars in eclipsing binary systems}
\subtitle{\targetnts: an optical blend\thanks{Based on observations made with the NASA/ESA Hubble Space Telescope, obtained at the Space Telescope Science Institute, which is operated by the Association of Universities for Research in Astronomy, Inc., under NASA contract NAS 5-26555. These observations are associated with program \#11223.}}

\author{A.~Pr\v sa\inst{1,2}
  \and E.~F.~Guinan\inst{1}
  \and E.~J.~Devinney\inst{1}
  \and S.~G.~Engle\inst{1,3}
}


\institute{Villanova University, Department of Astronomy, 800 Lancaster Ave, Villanova, PA 19085, USA
  \and University of Ljubljana, Dept.~of Physics, Jadranska 19, SI-1000 Ljubljana, EU
  \and James Cook University, Centre for Astronomy, Townsville QLD 4811, Australia
}

\date{Received xxx / Accepted xxx}

\abstract{During the OGLE-2 operation, Soszynski et al.~(2003) found 3 LMC candidates for an RR Lyr-type component in an eclipsing binary system. Two of those have orbital periods that are too short to be physically plausible and hence have to be optical blends. For the third, \targetnts, we developed a model of the binary that could host the observed RR Lyr star. After being granted HST/WFPC2 time, however, we were able to resolve 5 distinct sources within a 1.3'' region that is typical of OGLE resolution, proving that \target is also an optical blend. Moreover, the putative eclipsing binary signature found in the OGLE data does not seem to correspond to a physically plausible system; the source is likely another background RR Lyr star. There are still no RR Lyr stars discovered so far in an eclipsing binary system.}

\keywords{Methods: data analysis, observational -- binaries: close, eclipsing -- variables: RR Lyr -- stars: individual: \target -- techniques: astrometric, photometric}
\maketitle

\titlerunning{\targetnts: an optical blend}
\authorrunning{Pr\v sa, Guinan, Devinney, Engle}

\section{Introduction}

Eclipsing binary systems (EBs) have long been recognized as one of the most astrophysically rewarding targets for the reliable determination of masses, radii, luminosities and other physical stellar properties. The precision of the derived parameters is typically better than a few percent; it enables the studies of stellar structure and evolution, and yields reliable distances that are independent of any calibrations or empirical relations. EBs thus serve as \emph{astrophysical laboratories} suitable for the study of individual component properties. The importance of finding intrinsic variables in EBs is thus obvious: being able to obtain the fundamental properties of such stars provides improved theoretical understanding, better calibrations, and in some cases leads to improving the cosmological distance ladder \citep{guinan1998,fitzpatrick2003}. Cepheids, for example, have been found in galactic EBs \citep{freyhammer2005,antipin2007} as well as in the LMC \citep{alcock2002,lepischak2004,guinan2005}; there are more than 30 known $\delta$-Scuti type components in EBs \citep[see e.g.][]{dallaporta2002,rodriguez2004,christiansen2007}, and several other types such as slowly pulsating B stars \citep{pigulski2007,pilecki2007}. To date, though, no RR Lyr type star has been found in an EB. Recently, however, an accurate parallax and relative proper motion of RR Lyr itself was obtained with the HST Fine Guidance Sensor \citep{benedict2002} that estimated its absolute magnitude of $M = 0.61 \pm 0.1$.

The Optical Gravitational Lensing Experiment (OGLE) carried out a 4.5 square degree survey of the LMC during the second phase of operations (Jan 1997 -- Nov 2000; \citealt{soszynski2003}). They discovered 7612 RR Lyr-type objects: 5455 fundamental mode pulsators, 1655 first-overtone, 272 second-overtone, and 230 double-mode pulsators, along with several dozen other short-period pulsating variables. Three objects in their sample exhibited a superimposed RR Lyrae type variability on an eclipsing binary light curve: \targetnts, OGLE051822.60-691817.3, and OGLE050731.10-693010.3. These could be either the results of blending, or genuine RR Lyr stars in eclipsing binaries. In the latter case it would be possible, with additional spectroscopic data, to determine a reliable mass and radius estimate of an RR Lyr star for the first time. Given their (nearly) constant mean luminosities ($\langle L\rangle \sim 45L_\odot$ for RR Lyr type ab) and easily recognizable light curves, RR Lyr stars are important for the studies of the formation and evolution of population II stars, and for determining the galactic distance scale (globular clusters, LMC, and the local group; see \citealt{brown2004,sarajedini2006}). Although considerable effort went into calibrating the luminosity function for RR Lyr stars, it is compromised by dependences on metallicity, reddening, and possibly on period and/or pulsational modes \citep{sollima2006}.

\begin{figure*}
\centering
   \includegraphics[height=\textwidth,angle=-90]{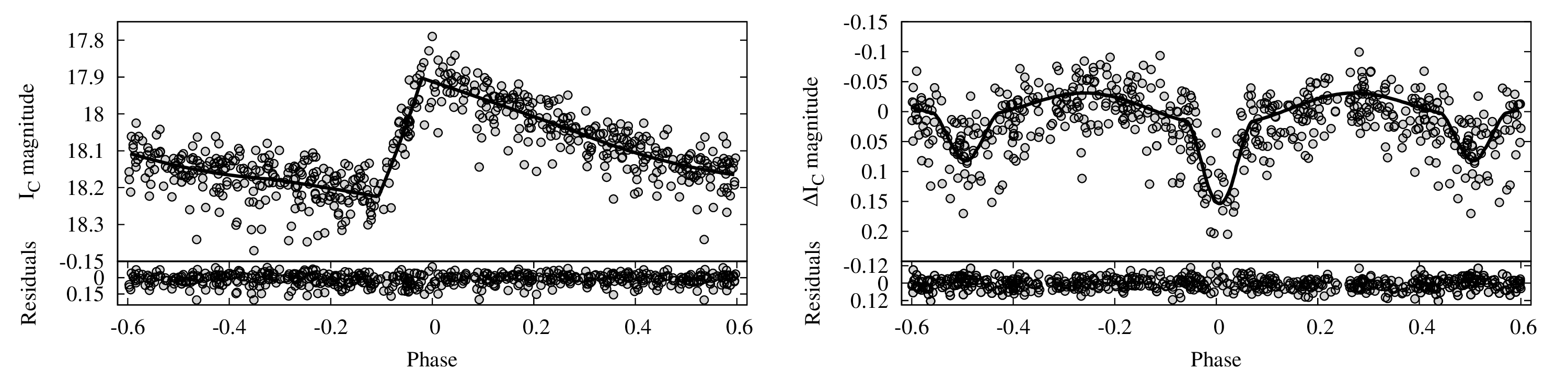}
     \caption{Left: light curve of \target phased with the RR Lyr pulsation period ($P_\mathrm{RRLyr} = 0.564876$-d). The solid line represents a 2nd order polynomial chain fit obtained by \emph{polyfit} \citep{prsa2008}. The apparent scatter under the curve is due to the putative EB signature. Right: residual light curve after subtracting the theoretical RR Lyr pulsation fit, phased at the EB period ($P_\mathrm{EB} = 8.92371$-d). The solid line depicts a model solution derived by {\tt PHOEBE}.}
     \label{lcs}
\end{figure*}

Of the three candidates, \citet{soszynski2003} concluded that, based on the period of the binary, the latter two are likely to be blends, leaving \target as the remaining candidate for an EB system hosting an RR Lyr component. We conducted a thorough feasibility study for all three candidates confirming their preliminary results: only the \target light curve could be plausibly attributed to an RR Lyr component in the EB system. The OGLE $I_C$ data (archive ID \verb|LMC_SC6_I_433519|) were passed through the period analysis algorithm PDM \citep{stellingwerf1978} and two distinct minima were detected: one pertaining to the RR Lyr ($P_\mathrm{RRLyr} = 0.564876$-d) and the other to the putative EB ($P_\mathrm{EB} = 8.92371$-d). The disentangling of the two contributions was done by \emph{polyfit} \citep{prsa2008} to the data phased at the $P_\mathrm{RRLyr}$ period (cf.~Fig.~\ref{lcs}, left). The residuals, folded at $P_\mathrm{EB}$, exhibit an EB-like shape (Fig.~\ref{lcs}, right). Assuming canonical values of parameters for the RR Lyr component: $\mathscr M_\mathrm{RRLyr} = 0.53 \mathscr M_\odot$ \citep{sandage2004} and $\mathscr R_\mathrm{RRLyr} = 5.4 \mathscr R_\odot$ \citep{marconi2005}, we were able to derive a solution consistent with the observed OGLE color of the system ($(B-V)_0 = 0.27$) and the expected absolute magnitude of an RR Lyr star ($M_V \sim 0.5$). To solve the light curve we used {\tt PHOEBE} \citep{prsa2005}, an analysis suite based on the Wilson-Devinney algorithm \citep{wd1971}. Our preliminary results indicated that the observed light curve could be plausibly explained by an EB hosting a horizontal branch (HB) primary and an RR Lyr secondary. Table \ref{params} lists the preliminary parameters obtained, which were used to compute the model light curve depicted in Fig.~\ref{lcs}.

\begin{table}
\caption{Preliminary parameters of \target HB--RR Lyr EB obtained by fitting the residual light curve depicted in Fig.~\ref{lcs} (right). Values denoted with an asterisk (${}^*$) are assumed. The errors in the table are \emph{formal}, derived from the covariance matrix.}
\label{params}
\centering
\begin{tabular}{lccc}
\hline\hline
Parameter:           &                 & System           &                   \\
                     & Primary (HB)    &                  & Secondary (RRLyr) \\
\hline
$a$ [$R_\odot$]      &                 & 18.5${}^\dagger$ &                   \\
$i$ [${}^\circ$]     &                 & $69.0 \pm 1.45$  &                   \\
$T_\mathrm{eff}$ [K] & $8300 \pm 113$  &                  & $6700 \pm 66$     \\
$\mathscr M$ [$\mathscr M_\odot$]      & 0.53${}^*$       &                 & 0.53${}^*$        \\
$\mathscr R$ [$\mathscr R_\odot$]      & $4.12 \pm 0.15$  &                 & $5.41 \pm 0.24$   \\
$M_\mathrm{bol}$     & $0.14 \pm 0.04$ &                  & $0.5^*$           \\
$L (I_C)$ []         & $6.36 \pm 0.09$ &                  & $6.44 \pm 0.08$   \\
\hline
\multicolumn{4}{l}{\rule{0pt}{2.6ex} ${}^\dagger$derived from $\mathscr M_1 + \mathscr M_2 = 1.06 \mathscr M_\odot$ and period $P=8.924$-d.}
\end{tabular}
\end{table}

\section{Observations}

To further investigate this system, we applied for multi-wavelength HST Wide Field Planetary Camera 2 (WFPC2) observations in Cycle 16 (PI Guinan, proposal ID \#11223). Table \ref{hstobs} summarizes the acquired observations and provides heliocentric Julian dates (HJD) and the orbital phases computed according to the following RR Lyr and EB ephemerides:
\begin{equation} \label{rrlyr_eph}
\Phi_\mathrm{RRLyr}^\textrm{\tiny max} = 2450452.24866 + E 0.564876,
\end{equation}
\begin{equation} \label{eb_eph}
\Phi_\mathrm{EB}^\textrm{\tiny min} = 2450452.466 + E 8.92371,
\end{equation}
where Eq.~(\ref{rrlyr_eph}) refers to the phase of maximum light and Eq.~(\ref{eb_eph}) refers to the EB primary minimum.

\begin{table}
\caption{HST/WFPC2 observation log. ID denotes HST archival identification. HJD values pertain to exposure starts. $\Phi_\mathrm{RRLyr} = (\mathrm{HJD}-2450452.24866)/0.564876$ and $\Phi_\mathrm{EB} = (\mathrm{HJD}-2450452.466)/8.92371$ are the phases of observations for RR Lyr and EB, respectively, where phase 0 for RR Lyr corresponds to maximum light, and phase 0 for EB corresponds to the primary eclipse (cf.~Fig.~\ref{lcs}).
}
\label{hstobs}
\centering
\begin{tabular}{rcccrr}
\hline\hline
ID: & Filter: & HJD$-$    & Exp: & \multicolumn{1}{c}{$\Phi_\mathrm{RRLyr}$} & \multicolumn{1}{c}{$\Phi_\mathrm{EB}$} \\
&        & $2450000$ & [s]  & & \\
\hline
1 & F255W   & 4298.94141 & 1000 & -0.1997 & 0.0399 \\
2 & F255W   & 4298.95460 & 1000 & -0.1764 & 0.0414 \\
3 & F555W   & 4298.96988 &  230 & -0.1493 & 0.0431 \\
4 & F300W   & 4299.00738 &  350 & -0.0829 & 0.0473 \\
5 & F300W   & 4299.01363 &  350 & -0.0719 & 0.0480 \\
6 & F336W   & 4299.02196 &  500 & -0.0571 & 0.0490 \\
7 & F555W   & 4299.03029 &  160 & -0.0424 & 0.0499 \\
8 & F380W   & 4299.03516 &  300 & -0.0337 & 0.0504 \\
9 & F439W   & 4299.07404 &  350 &  0.0351 & 0.0548 \\
a & F555W   & 4299.08099 &  160 &  0.0474 & 0.0556 \\
b & F675W   & 4299.08446 &  100 &  0.0535 & 0.0560 \\
c & F814W   & 4299.08863 &  160 &  0.0609 & 0.0564 \\
d & F953W   & 4299.09210 &  500 &  0.0671 & 0.0568 \\
e & F953W   & 4299.09974 &  500 &  0.0806 & 0.0577 \\
\hline
\end{tabular}
\end{table}

All exposures were corrected for geometric distortion, rotation, offsets, and scale differences between WFPC2 chips. For the purpose of analyzing the target (centered in the PC field), we extracted and trimmed the PC portion of the calibrated image. For the purpose of field astrometry and photometry we used all 4 fields corrected with IRAF's task {\tt metric}. Cosmic rays were cleaned using IRAF's {\tt crrej} task. In cases where a single exposure in a given filter was acquired (F336W, F380W, F439W, F675W, and F814W), the images were combined w.r.t.~the highest degree of similarity and with a 1\% scale noise to account for sub-pixel offsets. Photometry was done independently with IRAF's {\tt daophot} task \citep{stetson1987} and a stand-alone June 2008 version of the {\tt hstphot} program \citep{dolphin2000} that uses a PSF library to overcome the PSF undersampling issue. Due to the WFPC2 instrumental sensitivity drop in UV and IR there were insufficient counts in the F255W, F300W, F336W and F953W passbands to perform accurate photometry\footnote{Because of the HST/ACS failure in 2007 for which the initial proposal was made, the observations fell back on the less sensitive WFPC2 instrument. Moreover, the requested exposure times were computed for a single source of $I_C \sim 18$ that turned out to consist of 5 distinctly fainter sources.}. For the remaining passbands the results from both methods are within their standard errors.

\begin{figure}
\centering
   \includegraphics[width=\columnwidth]{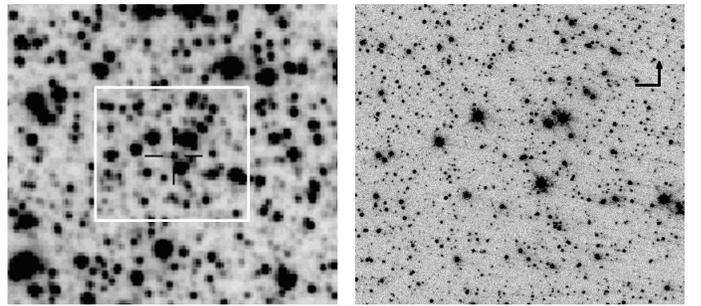}
     \caption{Left: the original OGLE finding chart with \target marked. North is up and east is to the left; the field is $60'' \times 60''$. Right: the observed PC field measuring $26.3'' \times 30''$ (marked with a white rectangle on the OGLE finding chart).}
     \label{finderchart}
\end{figure}

\section{Results}

\begin{figure}
\includegraphics[width=\columnwidth]{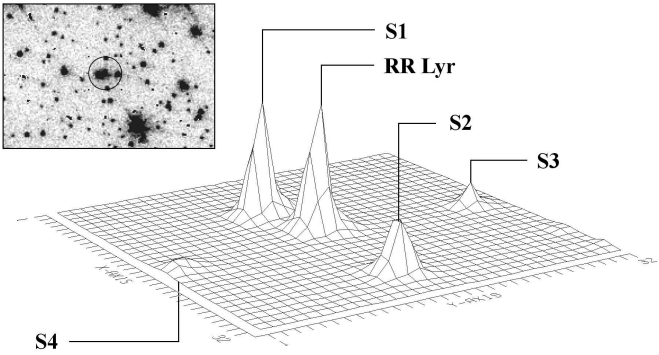}
\caption{Surface plot of the $1.34''$ field ($30 \times 30$\,px) in the vicinity of the target (ID3 exposure close to the RR Lyr minimum). The inset depicts the field with a circle of diameter 29.43\,px corresponding to the median OGLE resolution. The astrometric center of the image is a weighted photometric average. PC observations thus \emph{resolve} 5 distinct sources that contribute to OGLE photometry, indicating that \target is an optical blend rather than an RR Lyr star in an EB.}
\label{surface}
\end{figure}

The median seeing of OGLE photometry is estimated to $\sim$1.34$''$ \citep{soszynski2002}, which corresponds to $\sim$30 pixels in the PC field. Using high resolution HST/WFPC2 photometry, we were able to resolve as many as 5 distinct sources within the 30 pixel diameter centered on the target (cf.~Fig.~\ref{surface}) with their astrometric and photometric properties listed in Table \ref{results}. Two consecutive F555W exposures acquired during the minimum and maximum RR Lyr light (cf.~Table \ref{hstobs}) greatly facilitated its identification in the field. At $\sim$18.2$'$ from the optical center of the LMC, the estimated color excess of $E(B-V) = 0.075$ yields the maximum unreddened $(B-V)_0 = 0.112 \pm 0.05$, which corresponds to the maximum surface temperature of $T_\mathrm{max} = 8360\,\mathrm K \pm 400$\,K. The mean apparent magnitude $\langle V \rangle = 19.309 \pm 0.02$ is corrected for reddening ($A_V = 0.249$) and bolometric correction ($\Delta V = 0.06$); adopting the distance modulus of 18.43 to the LMC \citep{fitzpatrick2003}, this yields the absolute magnitude of the RR Lyr to be $\langle M \rangle = 0.57 \pm 0.02$, rendering \target a typical \emph{single} RR Lyr type ab.

\begin{table*}
\caption{Astrometric and photometric results for the sources in the vicinity of \targetnts. Right ascension $\alpha$ and declination $\delta$ are computed by flux-weighted averaging of the WCS data in all exposures where the sources were detected. The absolute astrometric accuracy of $\sim 0.1''$ is imposed by the Guide Star Catalog (GSC) accuracy, whereas the relative astrometric accuracy given in the table below does not exceed several mas. Photometry was performed using the June 2008 version of {\tt hstphot} for the aperture radius of $0.5''$ (11\,px) and gain 7. The values are given in the {\tt STMAG} system and are corrected for Charge Transfer Efficiency (CTE) and zero-point (ZP) offsets.
}
\label{results}
\centering
\begin{tabular}{llllllllll}
\hline\hline
Src: & \tcen{$\alpha$ [hms]} & \tcen{$\delta$ [dms]} & \tcen{$m_\mathrm{F555W}^\mathrm{\tiny ID=3}$} & \tcen{$m_\mathrm{F555W}^\mathrm{\tiny ID=7}$} & \tcen{$m_\mathrm{F380W}^\mathrm{\tiny ID=8}$} & \tcen{$m_\mathrm{F439W}^\mathrm{\tiny ID=9}$} & \tcen{$m_\mathrm{F555W}^\mathrm{\tiny ID=a}$} & \tcen{$m_\mathrm{F675W}^\mathrm{\tiny ID=b}$} & \tcen{$m_\mathrm{F814W}^\mathrm{\tiny ID=c}$} \\ [0.4ex]
\hline
RRL  & 5:22:17.9263 & -69:28:27.754 & 19.812( 9) & 18.889( 7) & 19.225( 12) & 18.993( 18) & 18.806( 7) & 18.719(11) & 18.459(  9) \\
S1      & 5:22:17.9720 & -69:28:27.839 & 19.887(12) & 19.874(14) & 20.316( 29) & 20.315( 37) & 19.884(16) & 19.645(30) & 19.241( 14) \\
S2  & 5:22:17.8252 & -69:28:27.779 & 20.654(14) & 20.588(17) & 20.940( 35) & 20.849( 37) & 20.566(17) & 20.466(28) & 20.174( 24) \\
S3  & 5:22:17.9155 & -69:28:27.147 & 21.826(27) & 21.782(34) & 22.099( 78) & 22.127( 74) & 21.773(34) & 21.685(62) & 21.192(187) \\
S4  & 5:22:17.9061 & -69:28:28.356 & 22.299(48) & 22.232(52) & 22.750(127) & 22.749(128) & 22.236(50) & 21.844(67) & 21.541( 66) \\
\hline
\end{tabular}
\end{table*}

The residual OGLE $I_C$ light curve depth of $\sim 0.15$ magnitudes implies that, with an average 3rd light contribution of $76.3\% \pm 0.5\%$ (64.8\% at RR Lyr minimum, 87.9\% at RR Lyr maximum), only the second brightest source (S1) could possibly be an EB, since the remaining sources could not cause the observed $\sim 15\%$ change in the composite light. The unreddened value of $(B-V)_0 = 0.355 \pm 0.05$ yields the effective temperature of the putative EB of $T_\mathrm{eff} = 6940\,\mathrm K \pm 245$\,K, making it an early-to-mid F type system. Given the fractional radius $\rho = R/a = 0.3$ that is determined by the eclipse width, such a binary would have an absolute magnitude of \emph{at least} $-0.5$, almost 2 magnitudes brighter than the distance to the LMC would allow. If, on the other hand, we imposed that the binary is in the LMC, the derived orbital semi-major axis of $\sim$12\,$R_\odot$ implies masses smaller than $0.2M_\odot$. In other words, there is no physically plausible model that would attribute the observed signature in the $I_C$ residuals to an EB.

The source is thus most likely a single star of mid-F spectral type and an unreddened absolute bolometric magnitude of $M = 1.12 \pm 0.02$. Plotting these values in a H-R diagram ($T_\mathrm{eff}=6940\,\mathrm K$, $L/L_\odot = 27$) places this star in the very center of the RR Lyr type region \citep[see, i.e., Fig.~1 of][]{gautschy1995}. To assess the possibility that the source is a shorter period pulsating star, we analyzed all weaker PDM minima and found the possible periods of $\sim $4.5-d (half-EB period) and $\sim 0.82$-d. The cadence of OGLE observations did not allow for a reliable search of even shorter periods. The shape of the residual light curve (i.e.~the absence of a saw-tooth-shaped signature), the 3rd light contaminated variability amplitude $\Delta I \sim 0.15$, and a larger absolute magnitude of $M \sim$1.12 indicates an RR Lyr type c star. These stars have periods between $0.25$-d and $0.4$-d \citep{szewczyk2008}, a period range unreachable with PDM using the OGLE data alone.

\begin{figure}
\includegraphics[height=\columnwidth,angle=-90]{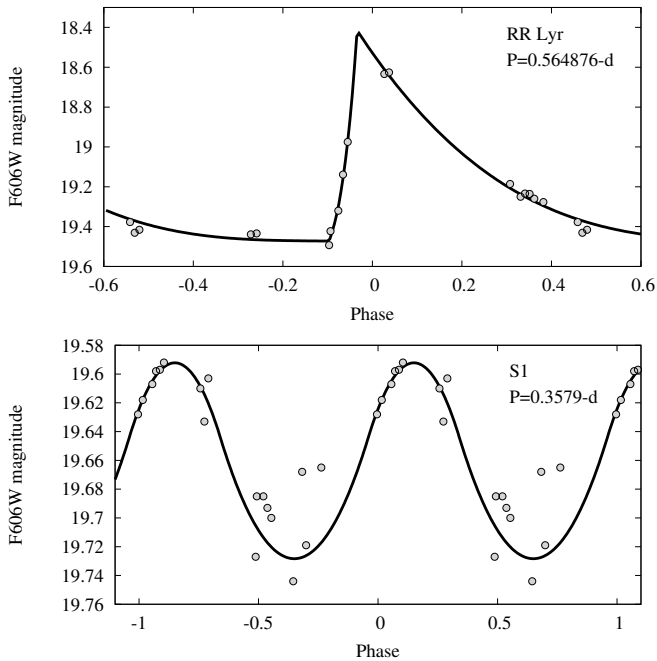}
\caption{Top: HST/ACS photometric curve of the RR Lyr folded according to the ephemeris given by Eq.~(\ref{rrlyr_eph}). Bottom: HST/ACS photometric curve of the source S1, folded to $P=0.3579$-d. Although this period has the lowest PDM $\theta$ value, it should not be considered significant due to a limited number of data points. Solid lines are computed by \emph{polyfit} and bear no physical significance -- they are included to help guide the eye.}
\label{acsphot}
\end{figure}

A field including \target had been observed previously in a broad V band (F606W) by HST/ACS in April and July 2006 (HST proposal ID 10753, PI Diaz-Miller), taking us a step closer to understanding the nature of this object. We reduced the ACS data by a similar approach as outlined in Section 2 and obtained an additional 18 photometric points listed in Table \ref{acsobs}. The data clearly show the star's variability with the amplitude of $\sim 0.18$ and a period of several hours (cf.~Fig.~\ref{acsphot}). There is insufficient information to attempt to determine the period (due to a very non-uniform time coverage), but the observations qualitatively support the RR Lyr type c hypothesis. Additional observations are needed to resolve the ambiguity. The remaining three sources (S2, S3, S4) are constant within their respective errors.

\begin{table}
\caption{HST/ACS photometry results for the RR Lyr and S1 sources. ID denotes HST archival identification. HJDs pertain to exposure starts.
}
\label{acsobs}
\centering
\begin{tabular}{lcccc}
\hline\hline
ID: & HJD$-$    & Exp: & V(RRLyr) & V(S1) \\
&   $2450000$ & [s]  & \\
\hline
j9it03ihq & 3851.61779 &  19 & 19.493(12) & 19.727(13) \\
j9it03iiq & 3851.61961 & 343 & 19.423(03) & 19.685(03) \\
j9it03ikq & 3851.62945 & 343 & 19.320(02) & 19.685(03) \\
j9it03imq & 3851.63539 & 343 & 19.139(03) & 19.693(03) \\
j9it03ioq & 3851.64131 & 343 & 18.975(02) & 19.700(03) \\
j9it03iqq & 3851.68745 & 343 & 18.634(02) & 19.668(03) \\
j9it03isq & 3851.69338 & 343 & 18.626(02) & 19.719(03) \\
j9it05sxq & 3854.68378 &  19 & 19.250(11) & 19.607(13) \\
j9it05syq & 3854.68934 & 343 & 19.234(02) & 19.598(03) \\
j9it05t0q & 3854.69520 & 343 & 19.236(02) & 19.597(03) \\
j9it05t2q & 3854.70114 & 343 & 19.260(02) & 19.592(03) \\
j9it05t4q & 3854.75582 & 343 & 19.377(03) & 19.610(03) \\
j9it05t6q & 3854.76173 & 343 & 19.431(03) & 19.633(03) \\
j9it05t8q & 3854.76767 & 343 & 19.416(03) & 19.603(03) \\
j9it06e6q & 3927.53951 & 423 & 19.186(02) & 19.744(03) \\
j9it06e8q & 3927.58158 & 423 & 19.277(02) & 19.665(03) \\
j9it07d5q & 3925.51787 & 423 & 19.439(02) & 19.628(03) \\
j9it07d7q & 3925.52501 & 423 & 19.434(02) & 19.618(03) \\
\hline
\end{tabular}
\end{table}

\section{Conclusions}

This paper establishes the nature of \target as a \emph{single} RR Lyr type star. Moreover, it shows the two-fold danger of third light contamination: 1) the binary model based on OGLE data alone provided a perfectly plausible physical description with virtually no means of resolving it without additional observations, and 2) although the residuals appeared to clearly point to the binary star signature, it seems that, somewhat ironically, the source of this secular variation is a distinct short period RR Lyr type c star. Being aware of these traps is all the more important in the era of fully automatic surveys and missions, where human supervision and a detailed object-by-object study is no longer possible. False positives present a serious challenge not only for ground-based observations, but for the upcoming space missions like Kepler and Gaia as well, emphasizing the importance of follow-up observations.

\begin{acknowledgements}
We gratefully acknowledge supporting grants HST-G0-11223.01A and NSF/RUI 05-07542. Image reduction and analysis were done with IRAF 2.14, STSDAS 3.8 and PyRAF 2.6. IRAF is distributed by the National Optical Astronomy Observatories, which are operated by the Association of Universities for Research in Astronomy, Inc., under cooperative agreement with the National Science Foundation. STSDAS and PyRAF are products of the Space Telescope Science Institute, which is operated by AURA for NASA.
\end{acknowledgements}

\bibliography{paper}

\end{document}